\documentclass[sigconf]{acmart}
\AtBeginDocument{%
  \providecommand\BibTeX{{%
    \normalfont B\kern-0.5em{\scshape i\kern-0.25em b}\kern-0.8em\TeX}}}

\copyrightyear{2024}
\acmYear{2024}
\setcopyright{acmlicensed}\acmConference[SIGCSE 2024]{Proceedings of the 55th ACM Technical Symposium on Computer Science Education V. 1}{March 20--23, 2024}{Portland, OR, USA}
\acmBooktitle{Proceedings of the 55th ACM Technical Symposium on Computer Science Education V. 1 (SIGCSE 2024), March 20--23, 2024, Portland, OR, USA}
\acmPrice{15.00}
\acmDOI{10.1145/3626252.3630855}
\acmISBN{979-8-4007-0423-9/24/03}

\begin{document}

\title[Failure Artifact Scenarios]{
Failure Artifact Scenarios to Understand High School Students’ Growth in Troubleshooting Physical Computing Projects
}

\author{Luis Morales-Navarro}
\email{luismn@upenn.edu}
\orcid{0000-0002-8777-2374}
\affiliation{%
  \institution{University of Pennsylvania}
  \city{Philadelphia}
  \state{Pennsylvania}
  \country{USA}
}

\author{Deborah A. Fields}
\email{deborah.fields@usu.edu}
\orcid{0000-0003-1627-9512}
\affiliation{%
  \institution{Utah State University}
  \city{Logan}
  \state{Utah}
  \country{USA}
}

\author{Deepali Barapatre}
\email{dee2496@upenn.edu}
\orcid{0009-0006-3505-3059}
\affiliation{%
  \institution{University of Pennsylvania}
  \city{Philadelphia}
  \state{Pennsylvania}
  \country{USA}
}

\author{Yasmin B. Kafai}
\email{kafai@upenn.edu}
\orcid{0000-0003-4018-0491}
\affiliation{%
  \institution{University of Pennsylvania}
  \city{Philadelphia}
  \state{Pennsylvania}
  \country{USA}
}

\renewcommand{\shortauthors}{Luis Morales-Navarro, Deborah A. Fields, Deepali Barapatre, \& Yasmin B. Kafai}

\begin{abstract}
  Debugging physical computing projects provides a rich context to understand cross-disciplinary problem solving that integrates multiple domains of computing and engineering. Yet understanding and assessing students’ learning of debugging remains a challenge, particularly in understudied areas such as physical computing, since finding and fixing hardware and software bugs is a deeply contextual practice. In this paper we draw on the rich history of clinical interviews to develop and pilot “failure artifact scenarios” in order to study changes in students’ approaches to debugging and troubleshooting electronic textiles (e-textiles). We applied this clinical interview protocol before and after an eight-week-long e-textiles unit. We analyzed pre/post clinical interviews from 18 students at four different schools. The analysis revealed that students improved in identifying bugs with greater specificity, and across domains, and in considering multiple causes for bugs. We  discuss implications for developing tools to assess students’ debugging abilities through contextualized debugging scenarios in physical computing. 
\end{abstract}




\maketitle

\section{Introduction}
Despite the fact that debugging is considered a key computing practice \cite{grover2013computational, lodi2021computational}, there are few instruments available to measure novice students’ thinking about debugging in K-12 contexts. Debugging is the process of troubleshooting in programming \cite{michaeli2021developing} that involves finding errors and fixing them \cite{mccauley2008debugging}. Research has shown that all students, but especially programming novices, face multiple challenges when debugging \cite{mccauley2008debugging} that range from identifying simple syntax problems to more complex semantic problems when programs run but do not function as intended. These challenges are even more evident in programming physical computing applications such as robots or electronic textiles (e-textiles) where debugging not only happens on-screen but also in the physical hardware \cite{desportes2019trials}. Debugging physical computing projects requires students not just to focus on the code that they have written to program sensors and actuators but also to attend to circuit design and physical construction issues \cite{searle2018debugging}. For this reason, debugging in physical computing is often called by the broader term troubleshooting \cite{jonassen2000toward}, which we use in the remainder of this paper when referring to identifying and solving problems in physical computing. To better understand novices’ learning to troubleshoot physical computing applications, we need instruments to capture and measure changes in how they engage with troubleshooting.

In this paper, we share findings from using a clinical interview protocol with two “failure artifact scenarios” with high school students before and after completing an e-textiles physical computing unit in which they learned to sew programmable circuits with conductive thread and connect sensors and actuators to a microcontroller. In addition, students participated in a debugging activity \cite{fields2021debugging} in which they created buggy e-textiles projects for their classmates to solve. The focus of this paper is on understanding changes in students’ approaches to troubleshooting, rather than on the unit itself. We conducted pre and post clinical interviews in which we described two “failure artifacts”—e-textiles projects that only partially worked, with accompanying images, creators’ intention statements for the projects and, for the second scenario, an interactive Scratch simulation of the project—and asked 18 high school students to give advice to the creators on how to identify and solve underlying problems.  In our analysis of the interview data we addressed the following research questions: How did students’ approaches to troubleshooting change from pre to post? In more detail, What types of bugs did students hypothesize? To what degree did students’ hypotheses and solutions account for the entirety of the multi-modality of the physical computing space? In the discussion we consider implications for developing tools to assess students’ debugging abilities through contextualized debugging challenges in physical computing.

\section{Background}

\subsection{Novice Challenges in Troubleshooting}

While  understanding novices' challenges with debugging has been of interest since  the early days of computing education research (e.g., \cite{spohrer1986novice}), relatively few efforts have focused on assessing students’ approaches to debugging in K-12. Indeed, a systematic mapping on assessment of computational thinking found only two studies in which debugging was assessed \cite{de2016systematic} and a more recent systematic review identified only one study in which debugging was evaluated \cite{tang2020assessing}. In one of these studies pre-service teachers were presented with a troubleshooting scenario (i.e., a lamp not working) and asked to select one possible step to fix the lamp among five choices \cite{yadav2014computational}. Another study used open-ended survey questions to gather students’ perspectives on using debugging tools in a game-based programming environment \cite{kazimoglu2012learning}. The third assessment discussed in the systematic reviews involved using rubrics to observe young children and evaluate how they debugged simple programs to move a robot \cite{bers2014computational}.

Assessing troubleshooting is particularly important as novices encounter difficulties in finding and fixing bugs \cite{mccauley2008debugging}. In assessments we must consider that troubleshooting involves constructing a problem space, isolating or diagnosing issues, and finally generating solutions \cite{jonassen2006learning,michaeli2021developing}. Constructing a problem space requires learners to build a mental model to represent the system they are troubleshooting and how its parts are interconnected. Isolating issues or diagnosing them involves observing problems and ascribing plausible causes. Here learners may rely on their previous experiences with issues they have encountered in the past \cite{konradt1995strategies}. Diagnosing issues often involves generating hypotheses and testing them at different levels, one may test the whole system or focus on only a part of it \cite{jonassen2006learning}. Lastly, generating solutions involves coming up with plausible ways to address an identified issue. 

More recent work has investigated students’ “debugging traits”, identifying troubleshooting-like practices that even absolute novices may have already developed. Here, the work of Michaeli and Romeike \cite{michaeli2021developing} assessing troubleshooting highlights the need for better instruments to evaluate how novices troubleshoot the problems they encounter. They developed an instrument to assess troubleshooting through escape room tasks. With this instrument they found that novice students encountered problems formulating initial hypotheses, coming up with alternative hypotheses or with multiple hypotheses of what the cause of a problem could be \cite{michaeli2021developing}. This is not surprising since being able to generate plausible hypotheses and multiple hypotheses for a single issue is a documented difference between novice and experienced troubleshooters \cite{gugerty1986debugging, kim2018debugging}. 

\subsection{ Failure Artifacts for Assessments}
In this paper we build on a longstanding instructional practice in computing education of providing learners with what we broadly call “failure artifacts,” i.e., buggy code or buggy physical computing projects. This practice dates back to the early days of computing education research. For instance, Carver and Klahr \cite{carver1986assessing} provided students with buggy Logo programs, and asked them to identify and solve code problems. Schwartz, and colleagues \cite{schwartz1989metacourse} followed with the development of what they called a “Metacourse” in which they trained and asked students to identify and then fix different types of bugs in given Basic programs. Harel \cite{harel1990children} and Kafai \cite{kafai2012minds} also provided students with buggy Logo programs to fix, both on paper and on the computer.

Less attention has been given to troubleshooting physical computing projects, which involves not only debugging code but also fixing issues in circuitry. Physical computing requires learners to understand how different hardware and software components interact \cite{desportes2019trials, wagh2017role} and how bugs may emerge across and within domains such as coding, circuitry and craft \cite{searle2018debugging}. Working with both hardware and software may make troubleshooting more difficult, as DesPortes and DiSalvo \cite{desportes2019trials} demonstrate in studies of learners creating Arduino projects. They found that novices frequently identify bug locations incorrectly (e.g., identify bugs in only one domain, such as circuitry), ignore errors in one domain (e.g., code), or incorrectly assume bugs are solved. As such, being able to identify plausible bugs, generate multiple hypotheses, and understand how to troubleshoot across domains (i.e., circuitry and code) is essential for learning physical computing.

Beyond just code, some studies have given buggy physical computing projects, such as e-textiles,  to students in order to support learning troubleshooting and research student thought processes while they solve projects peppered with multiple problems \cite{fields2016deconstruction, jayathirtha2020pair}. This work highlights how assessing troubleshooting in physical computing can be particularly difficult for teachers as students often work to create open-ended, personally relevant projects \cite{przybylla2017nature}. Research in this area has shown that in physical computing, its multi-modal nature (with bugs in code and circuitry), complicated by three-dimensional spaces where circuits traverse the front, back, and insides of artifacts generate challenges for students to identify bugs. Notably, many of these studies applied clinical interviews or “think aloud” interviews to elicit students’ thought processes as they hypothesize, test, narrow down, identify, and eventually solve bugs distributed throughout multi-modal physical computing projects \cite{jayathirtha2020pair}. However one challenge to such projects is the length of time it takes for students to identify and solve multiple bugs in a multi-modal computing system.

During clinical interviews learners are presented with problematic situations and asked to solve them, explain them or think about them \cite{disessa2007interactional}. Here interviewers encourage learners to verbalize their thinking processes and explain their thinking, seeking to uncover student understandings \cite{disessa2007interactional}. These types of interviews can also serve as assessment tools for teachers and researchers to better understand students’ knowledge and experiences in specific domains and tailor their instruction according to students’ understanding to build on their current knowledge \cite{russ2013using}. In computing, clinical interviews have been used with undergraduate students to investigate how they construct knowledge around difficult concepts in CS1 \cite{yuen2007novices} and how they reason through how commercially available physical computing artifacts work before and after a physical computing class \cite{lee2017rubric}. In K-12 computing education, interviews have been occasionally used to assess programming abilities \cite{sentance2023formative} affording researchers opportunities to prompt learners to demonstrate their understanding of concepts \cite{grover2015designing}, and interact with design scenarios to fix bugs and remix projects \cite{brennan2012new}.

In this study, we developed a clinical interview protocol that draws on the tradition of presenting students with a buggy project to solve: a failure artifact scenario. However, instead of giving students an actual physical computing project and its code which would require substantial time to investigate and solve, we provided students with descriptions of two buggy e-textiles projects that did not work as intended. The goal of the interview was to elicit how students would go about troubleshooting, what suggestions they would give to the creators (i.e., imaginary students in their class) and to what degree they would provide insights into their thought processes of troubleshooting. Analysis allowed us to look at changes in students’ approaches to troubleshooting.

\begin{figure*}[h]
  \centering
  \includegraphics[scale=0.7, width=\linewidth]{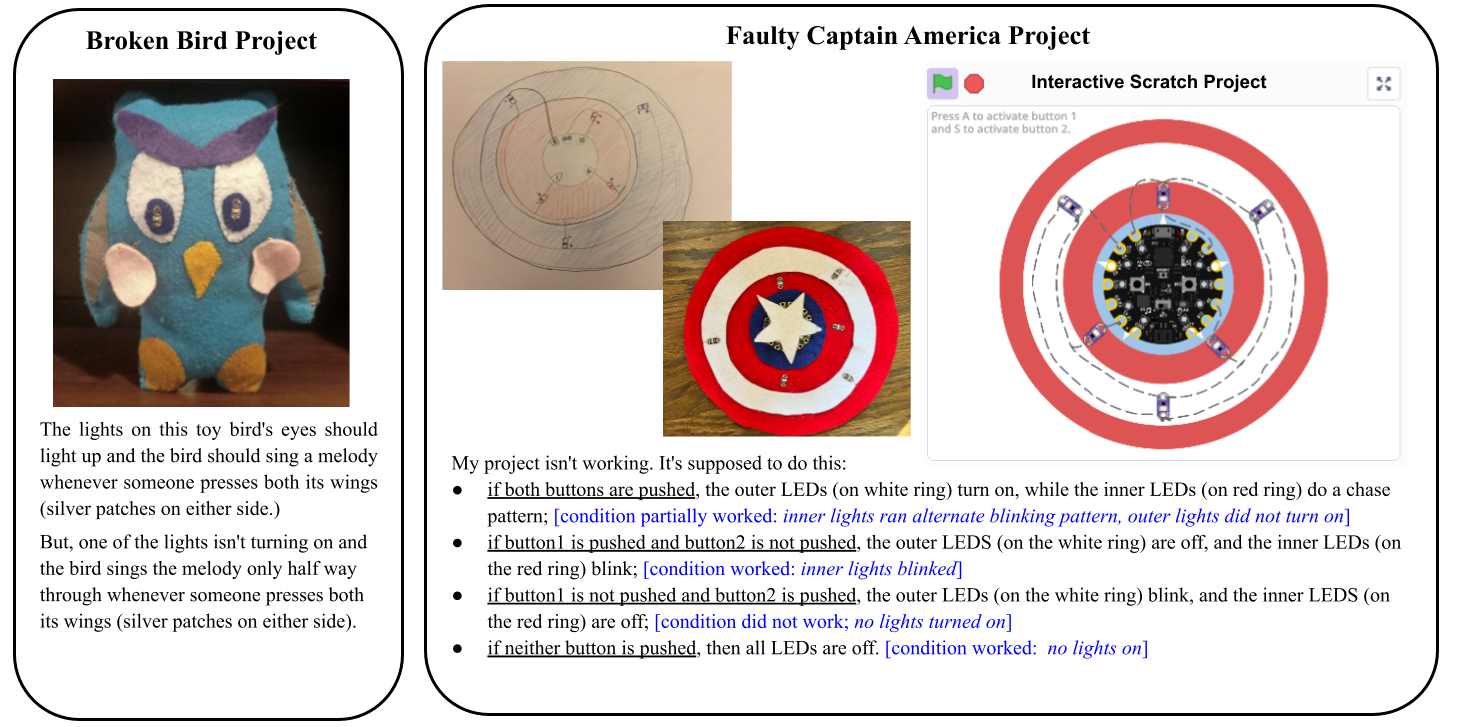}
  \caption{Interview Prompts Featuring Broken Projects. Comments in blue indicate actual functionality of interactive Scratch project.}
  \Description{Interview prompts}
\end{figure*} 

\section{Methods}   

\subsection{Context}
Our interviews were scheduled before and after students participated in the e-textiles unit of the \emph{Exploring Computer Science} (ECS) course, a year-long, equity-focused, inquiry-based introductory computing course for secondary students that has been adopted by school districts across the United States \cite{goode2012beyond}. The 10-12 week \href{http://exploringcs.org/e-textiles}{e-textiles unit} is designed for youth to create a series of personally relevant creative projects while learning new coding, circuitry, and crafting technical skills \cite{kafai2019stitching}. The unit requires students to apply computing concepts such as sequences, loops, conditionals, variables, nested conditionals, data input from sensors, and functions in a text-based programming language (Arduino). Included in this study between the third and fourth projects of the e-textiles unit was an additional 7-day debugging activity, where students created intentionally buggy projects for their peers to solve (for more on the activity see \cite{fields2021debugging, morales2021growing}). In total, students spent 8-12 weeks creating, coding, and debugging e-textile projects. 

Of note, all activities of this study—learning and research—took place virtually through Zoom in Spring 2021 because all schools in the area had virtual schooling due to COVID-19. Students were provided materials through pick-up at school or directly through the mail, depending on each school’s circumstances. The study protocol was approved by the University of Pennsylvania's institutional review board.

In Spring 2021, four experienced teachers at different secondary schools in the Western United States, with high percentages of historically marginalized secondary student populations (58-95\% free and reduced lunch; 85-99\% non-white), taught eight ECS classes with e-textiles. Two researchers who participated in the classes conducted pre-interviews with 33 consenting students. Students were selected randomly from those who provided informed consent (parents) and assent (students), with ~4-5  students per class. Because of the timing of the study at the end of the virtual school year in the second year of the pandemic, there was a high degree of attrition in student participation; many students were no longer participating by the end of the school year. In the end, we had 18 matching pre- and post-interviews.

\subsection{Data Collection}

Data were collected through semi-structured clinical interviews. These interviews involved a set of two predetermined failure artifact scenarios with prompts designed to solicit student reasoning in an open-ended manner. We intentionally designed these interviews to be similar to those used in conceptual change research \cite{disessa2007interactional, lee2017rubric, sherin2012some}. The appeal of open-ended conceptual change style interviewing was that they could be accessible to novices, elicit a broad range of student ideas (e.g., about debugging, e-textiles, and computing), and followed-up with questioning for increased robustness. Interviews took an average of 10.5 minutes each lasting between 5 - 25 minutes.

The interview protocol we developed invited students to make suggestions for how to debug two projects that someone else had made. Images of the broken projects and descriptions of how they were supposed to work are shown in Figure 1. The interviewer read the descriptions outloud along with the written text made available to students then asked students for help:  “What would you tell this student to do in order to fix the project?”

For the second project, the Captain America shield, we also presented students with an interactive \href{https://scratch.mit.edu/projects/479329018}{Scratch project} that enabled them to test how the project was not working. If students carefully tested all conditions, they would notice that the outer lights did not work under any condition, but that pressing button2 and button1 at the same time resulted in a different pattern for the inner lights than just pressing button1. This was important since it meant that nothing was wrong with button2 as a mechanism - pressing it triggered a new condition. Thus the second scenario became a means of assessing how systematically students tested the project and whether they could eliminate some possible problems through that testing. 

In the post-interviews, the scenarios remained almost identical but with slightly different looking projects (e.g., an elephant and a teddy bear with identical circuitry and functioning).

In addition to the general prompt (to help a peer fix the projects), when students ran out of ideas we used follow-up prompts: “What do you think could be the causes of each of these issues? How would you fix them?” This allowed us to note whether students’ ideas came up without or with prompting.

\subsection{Analysis}

We analyzed 18 pre and 18 post interviews in two rounds, with a total of 4 hours and 51 minutes of video recordings. Interviews were transcribed, and transcriptions checked for accuracy. In our analysis we build on traditions of qualitative and learning sciences research in computer science education \cite{tenenberg2019qualitative,margulieux2019learning}. During a first round of analysis two researchers inductively coded a third of the data (6 sets of matching pre/post interviews) to develop an initial coding scheme. We checked and revised this scheme several times in group meetings and with application to additional interviews, then created a codebook with categories for identification of bugs, multiple causes, next steps for fixing bugs, debugging process, testing, and whether students made references to their personal experiences in relation to debugging.  Following, in a second round of analysis two researchers applied the coding scheme across all pre/post interviews, co-watching the video recordings together while also notating the transcript to capture gestures and participant testing of the interactive Scratch simulation of the broken project. While coding together the researchers dialogically engaged with the data, seeking agreement and iteratively discussing disagreements with a third researcher familiar with the data and the coding scheme. Since this is an exploratory study with a small number of participants, we prioritized establishing unanimous agreement on all coding (by coding together and reconciling our different opinions through extensive discussion) over reliability (when coders apply the same scheme independently on the same data) \cite{mcdonald2019reliability}. All names used in the paper are pseudonyms. 

\section{Findings}

Prompting students with failure artifact scenarios during the interviews enabled us to observe growth in thinking about troubleshooting e-textiles. In particular, in post more students were able to identify potential bugs across domains, they identified multiple causes for bugs more frequently, and identified more specific bugs.

\subsection{Identifying potential bugs across domains}

First, from pre to post students improved their identification of  bugs across \emph{multiple domains}: code, circuitry and both domains (see Figure 2).  This is important in the case of physical computing where a bug may be caused due to code (e.g., a variable to store an input pin number not being declared) and/or circuitry issues (e.g., negative and positive touching and causing a short circuit). Considering multiple domains in identifying bugs  shows that students construct a complex problem space that involves both code, circuitry, and the interactions between domains. Whereas prior to the intervention 8 students (44\%) identified both circuit and coding bugs, afterwards 13 students (72\%) identified bugs in both domains. After the intervention more students were aware that bugs in physical computing systems may be present both in circuitry and code.

\begin{figure}[h]
  \centering
  \includegraphics[scale=0.35]{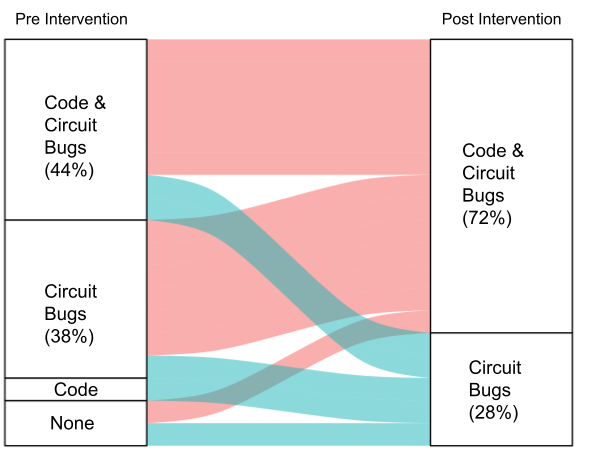}
  \caption{Alluvial diagram shows the distribution of students by domain type of bugs identified.}
  \Description{Alluvial diagram.}
\end{figure}

\subsection{Identifying multiple causes for potential bugs}

\begin{figure}[h]
  \centering
  \includegraphics[scale=0.35]{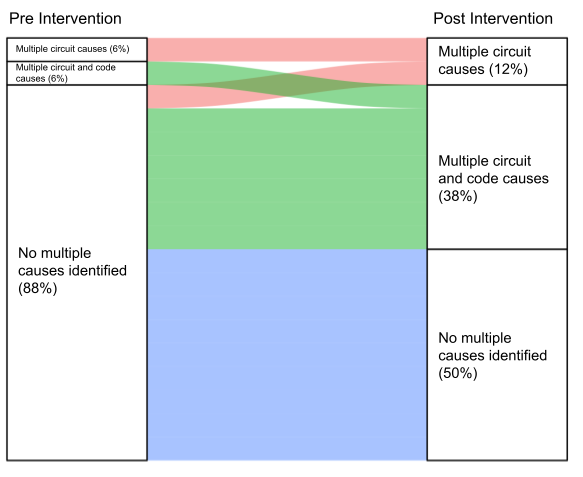}
  \caption{Alluvial diagram showing distribution of students that identified that a bug may be caused by more than one problem in a single domain or across domains.}
  \Description{Alluvial diagram.}
\end{figure}

Second, students showed pronounced differences between pre and post in identifying multiple causes for bugs within single domains and across both circuitry and code (see Figure 3). For instance, an LED light could fail to turn on because of a short circuit or because the pin that connects it to the microcontroller was not correctly declared in the code. Overall, while in pre only two students (12\%) voiced that a bug could be caused by more than one issue, in post 9 students (50\%) identified multiple causes for bugs. For instance, Ava went from not identifying any possible bugs in pre to identifying that a light may not turn on because of two possible causes \emph{across domains}: a faulty connection or not correctly declaring a variable that stores the number of the pin where the LED light is connected. 

Further, of the two students in pre who identified that a bug may be caused by multiple issues, only one student identified that bugs could be caused by \emph{multiple issues across domains}. One student, Amelia, identified \emph{multiple circuitry causes} for a bug, saying that it would be possible that an LED did not turn on because the component was broken or due to a loose connection. Fernando was the only student that identified \emph{multiple causes in circuitry and code} in pre, voicing that the music could be failing to play because of issues in circuitry (a short circuit) or code (a function in the code was not set to play the whole song).

In contrast, in post 7 (38 \%) students identified that individual bugs—such as lights not turning on, a motor not moving, a button not working, lights not blinking— could be caused by multiple issues across domains in coding or circuitry. For example, Damian explained that an LED could fail because the creator of the project did not declare the variable for the pin correctly, because they did not pass the right arguments to the \texttt{digitalWrite()} function or because of a loose connection to the circuit board.

Generating multiple hypotheses in debugging is a documented difference between novices and more experienced students \cite{michaeli2020investigating, kim2018debugging, gugerty1986debugging}. It is notable  that in post interviews, more students identified multiple causes for bugs, and oftentimes multiple causes across different domains, showing their increasing understanding of the problem situation and the interconnected domains of physical computing.

\subsection{Becoming more specific in identifying bugs}
Third, students not only identified more possible coding bugs but also moved towards \emph{more clearly defining} possible bugs after the e-textiles unit. Students that identified possible coding bugs came up with 1.89 bugs on average (SD = 1.05) in the pre-interview and 3.85 bugs on average (SD = 3.11) in the post-interview. We classified the bugs identified into three categories defined, semi-defined, and undefined bugs (see Table 1). 

\begin{table}[]
\caption{Coding bugs identification categories with definitions and examples.}
\scalebox{0.75}{
\begin{tabular}{l|l|l}
\hline
\textbf{\begin{tabular}[c]{@{}l@{}}Bug \\ Identification \\ Categories\end{tabular}} & \textbf{\begin{tabular}[c]{@{}l@{}}Description: Identify coding \\ as the relevant domain with \\ varying levels of definition\end{tabular}} & \textbf{Example}                                                                                                                 \\ \hline
\textbf{\begin{tabular}[c]{@{}l@{}}Defined \\ Coding Bugs\end{tabular}}              & \begin{tabular}[c]{@{}l@{}}Identify a bug location, and \\ identify a specific error.\end{tabular}                                           & \begin{tabular}[c]{@{}l@{}}“Using digitalRead() instead \\ of digitalWrite() to change \\ that LED to HIGH.” Damian\end{tabular} \\ \hline
\textbf{\begin{tabular}[c]{@{}l@{}}Semi-defined \\ Coding Bugs\end{tabular}}         & \begin{tabular}[c]{@{}l@{}}Identify either a bug location\\  or a specific error, but not both.\end{tabular}                                 & \begin{tabular}[c]{@{}l@{}}“There’s something wrong \\ in the for loop”  Viviana\end{tabular}                                    \\ \hline
\textbf{\begin{tabular}[c]{@{}l@{}}Undefined \\ Coding bugs\end{tabular}}            & \begin{tabular}[c]{@{}l@{}}No bug locations or \\ specific errors identified.\end{tabular}                                                   & \begin{tabular}[c]{@{}l@{}}“Since there's two it's, \\ maybe one of the lights \\ isn't coded in yet” Amelia\end{tabular}        \\ \hline
\end{tabular}
}
\end{table}

\begin{figure}[h]
  \centering
  \includegraphics[scale=0.28]{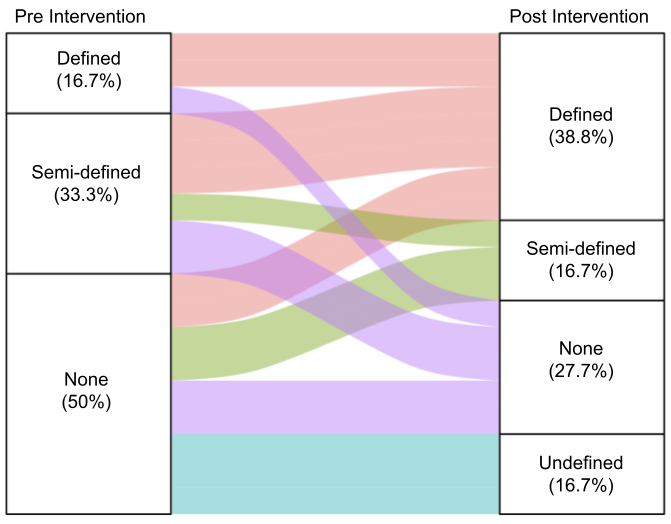}
  \caption{Alluvial diagram showing how student distribution by coding bugs identification category changed.}
  \Description{Alluvial diagram.}
\end{figure}

Overall, students improved over time in defining coding bugs more clearly, moving from undefined to semi-defined coding bugs and from semi-defined to defined coding bugs. The number of students that came up with possible defined bugs (specific bugs in specific locations) increased from 3 students (16\%) in pre to 7 students (38\%) in post. Figure 4 shows how 3 students moved from identifying semi-defined bugs to defined bugs and 2 students from not identifying bugs to identifying defined bugs. Fernando, for example, moved from identifying only one semi-defined bug (i.e., something wrong in the code when the button is pressed) to suggest defined bugs such as not updating the variable used to store the value of \texttt{digitalRead(“button1”)} prior to using it in a conditional statement. Here, not only the number of students who correctly identified defined bugs increased, but the depth of understanding of computing in how they described possible bugs also varied. For example, while in pre Cameron suggested that perhaps the creator of the project only included half of the notes for the song, a defined bug that would prevent the project from working as expected, in post he suggested that perhaps the creator of the project made a mistake in the conditional statements that check for the status of a button setting the condition to \texttt{LOW} instead of \texttt{HIGH}. This last bug shows a deeper understanding of computing concepts.

Observing the level of definition in which students came up with possible bugs is helpful to capture qualitative changes in how they diagnosed bugs and generated hypotheses. Consider explanations that identified undefined bugs pointing to errors in the code without providing any details on what the bugs could be or where these could be located. Three students moved from not identifying any bugs to identifying undefined bugs. Examples of undefined bugs include Ernesto saying that “maybe something's wrong with the code” or Karina explaining that “maybe in the code, she didn't write it right, or she like missed a step, that's why one of the LED light is not on”. These explanations show vague ideas about how coding might be a problem area but without much clarity outside of missing a step.

Contrast the undefined bugs with semi-defined bugs that included, for example, suggesting that “maybe you have to fix like an input” identifying that there could be a bug in how the buttons are declared as inputs without explaining what the specific bug could be. Another example is “there’s something wrong in the for loop,” providing a location for a bug but not a specific bug. These instances show that even though students may not have a concrete idea of what bugs may be causing an issue, they had some understanding of where in the code to look for bugs. On the other hand, some issues were identified without determining their location. In post, Salim explained that there could be something wrong with the \texttt{HIGH} and \texttt{LOW} without providing details on whether he thought these issues were in the conditional statements or in the \texttt{digitalWrite()} functions used to turn on the LEDS. Looking for changes in students’ level of definition in identifying bugs in code was a productive way to see improvement in troubleshooting.

\section{Discussion}
Using failure artifacts scenarios in the interviews revealed students’ growth in troubleshooting e-textile projects: (a) students became competent in identifying bugs across domains, (b) they improved in identifying multiple causes for potential bugs and (c) they became more specific in articulating what was wrong. These are valuable computing competencies in learning how to design and fix physical computing projects. Being able to capture how students identify bugs across domains is particularly important since physical computing novices often identify bug locations incorrectly, assuming errors occur mostly in circuitry \cite{desportes2019trials}. At the same time, observing students' capacity to identify multiple causes for potential bugs across domains is key as this is a documented difference between novice and more experienced debuggers \cite{kim2018debugging,michaeli2020investigating}. Finally, seeing how students became more specific about the bugs they proposed may show growth in their understanding of the problem space, as well a wider repertoire of previous troubleshooting experiences to draw upon.

Further, with the introduction of failure artifact scenarios in clinical interviews to assess debugging, we present physical computing instructors with an approach to engage students, individually or collaboratively, in demonstrating their newly gained competencies. In our particular case, we designed the failure artifacts scenarios in intentional ways based on our extensive knowledge of the challenges novices encounter when creating physical computing artifacts. The problems were explicitly vague (i.e., the toy only plays half of a song and half of the lights do not work) to elicit as many troubleshooting strategies and thinking processes as possible without the intensive labor of actually solving a project. Thus the failure artifact scenarios clinical interview provided a less time consuming means of inquiring into students’ troubleshooting processes while still providing insights into their abilities to think across domains of physical computing, identify multiple potential causes for bugs, and provide explicit ideas about potential bugs. 

From this study it is also possible to draw implications for assessing students’ learning in physical computing. While being able to design functional circuits or write code are critical, it is at the intersection of these two domains that the ability to create functional computational artifacts lies. As Russ and Sherin \cite{russ2013using} demonstrate, these types of interview scenarios can serve as assessment tools for teachers to better understand students’ knowledge. Future research could investigate how teachers may adopt failure artifact scenarios as assessments tools in the classroom. 

As a research tool, the failure artifact scenarios may provide a starting point for others to assess and study students’ understanding of troubleshooting. Future studies could design failure artifact scenarios in other domains of physical computing such as robotics. They could also be applied in a quasi-experimental design to see if one intervention was more successful at supporting students’ learning of troubleshooting than a control group. For this it might be helpful to develop the failure artifact scenario into a survey construct, exploring in a pre- post- measure whether students identify multiple causes of problems, across multiple domains, and with increasing levels of specificity.  

Our findings provide a first step towards understanding what failure artifact scenario interviews can reveal about students’ thinking on debugging and troubleshooting. We look forward to future research that explores other failure artifact scenarios, expands our understanding of how novices learn to troubleshoot, and contributes to a more robust collection of tools to assess debugging in K-12 physical computing.

\begin{acks}
With regards to Katherine Gregory for support in data analysis. This work was supported by a grant from the National Science Foundation to Yasmin Kafai (\#1742140). Any opinions, findings and conclusions or recommendations expressed in this paper are those of the authors and do not necessarily reflect the views of NSF, the University of Pennsylvania or Utah State University.
\end{acks}

\bibliographystyle{ACM-Reference-Format}
\balance
\bibliography{sample-base}

\end{document}